\documentclass{sigplanconf}
\usepackage{etex}

\usepackage{xspace}
\usepackage{xcolor}
\usepackage{amsmath,amssymb,amsthm}
\usepackage{hyperref}

\usepackage{logic}

\usepackage{bussproofs}
\usepackage{proof}
\EnableBpAbbreviations

\usepackage{fancybox}

\newtheorem{lemma}{Lemma}
\newtheorem{theorem}{Theorem}
\newtheorem{corollary}{Corollary}

\theoremstyle{definition}

\newtheorem{example}{Example}

\theoremstyle{remark}

\usepackage{mathtools}

\newenvironment{calculus}
{\begin{center}\begin{Sbox}\begin{minipage}[b]{0.46\textwidth}}
{\end{minipage}\end{Sbox}\fbox{\TheSbox}\end{center}}

\newcommand{\UPRCL}{\textbf{CR}\xspace}

\begin{document}

\setlength{\pdfpageheight}{\paperheight}
\setlength{\pdfpagewidth}{\paperwidth}

\conferenceinfo{CONF 'yy}{Month d--d, 20yy, City, ST, Country}
\copyrightyear{20yy}
\copyrightdata{978-1-nnnn-nnnn-n/yy/mm}
\copyrightdoi{nnnnnnn.nnnnnnn}

% Uncomment the publication rights you want to use.
%\publicationrights{transferred}
%\publicationrights{licensed}     % this is the default
%\publicationrights{author-pays}

\titlebanner{Draft}        % These are ignored unless
\preprintfooter{First-Order Unit-Propagating Resolution with Clause Learning}   % 'preprint' option specified.

% Page limit: 10

\title{Conflict Resolution}
\subtitle{a First-Order Resolution Calculus with \\ Decision Literals and Conflict-Driven Clause Learning}

\authorinfo{John Slaney \and Bruno Woltzenlogel Paleo}%\and Name3}
           {Australian National University}
           {john.slaney@anu.edu.au \quad bruno.wp@gmail.com}

\maketitle

\begin{abstract}
This paper defines the (first-order) conflict resolution calculus: an extension of the resolution calculus inspired by techniques used in modern SAT-solvers. The resolution inference is restricted to (first-order) unit-propagation and the calculus is extended with a mechanism for assuming \emph{decision literals} and a new inference rule for \emph{clause learning}, which is a first-order generalization of the propositional \emph{conflict-driven clause learning} (CDCL) procedure. The calculus is sound (because it can be simulated by natural deduction) and refutationally complete (because it can simulate resolution), and these facts are proven in detail here.
\end{abstract}

\category{F.4.1, I.2.3}{mathematical logic, deduction and theorem proving}{proof theory, deduction}

% general terms are not compulsory anymore,
% you may leave them out
% \terms
% term1, term2


\section{Introduction} 

Modern SAT-solvers are famously efficient for solving the decision problem of
satisfiability of propositional formulas, and we may
wonder whether the ideas used in SAT-solvers could be generalized to
the first-order case. This paper addresses this question from a purely
proof-theoretical perspective.

We briefly recall the first-order resolution calculus (in Section
\ref{sec:Resolution}), which is the theoretical foundation for many
current state-of-the-art first-order theorem provers (e.g. \cite{Vampire,E,SPASS}), and
the DPLL and CDCL procedures used by SAT-solvers (in Section \ref{sec:DPLLCDCL}).
The main contribution of this paper (presented in Section
\ref{sec:UPRCL}) is the \emph{conflict resolution} calculus \UPRCL. 
It extends the first-order resolution calculus with 
\emph{decision literals} and a new inference rule for
\emph{clause learning} and restricts the
resolution rule in order to force it to behave like \emph{unit propagation}. 
As discussed in Subsection \ref{sec:Isomorphism}, a certain subclass of \UPRCL derivations 
is isomorphic to the abstract data structure 
known as \emph{conflict graphs} or \emph{implication graphs} and widely used to describe the procedures of modern SAT-solvers.
Furthermore, as shown in Section \ref{sec:Splitting}, whereas the \emph{splitting} technique used by modern first-order provers must either be handled at an extra-logical level or lead to an unacceptable increase in proof size if simulated in the resolution calculus, its simulation by \UPRCL's decisions and clause learning is lean and straightforward. Therefore, the new \UPRCL calculus provides a more adequate proof-theoretical foundation for procedures currently implemented by SAT-solvers and first-order provers.

In \UPRCL, it becomes evident that decision literals are
analogous to assumptions in natural deduction, whereas clause learning
resembles natural deduction's implication introduction rule. This fact
is crucial for the proof of soundness of \UPRCL (shown
in Section \ref{sec:Soundness}) and it illustrates an insightful novelty
of the calculus: while the resolution inference proposed by Robinson
(\citeyear{Robinson}) can be regarded as a first-order generalization of
modus ponens (a.k.a. natural deduction's implication
\emph{elimination}) by taking unification into account, the clause learning
rule proposed here (and inspired by the propositional CDCL technique)
can be considered a first-order generalization of implication
\emph{introduction}, as it discharges decision literals in a way that allows for unification. 

Any resolution refutation can be translated into a refutation in the
proposed calculus. Therefore, \UPRCL's refutational completeness follows easily 
from the refutational completeness of the resolution calculus (as
demonstrated in Section \ref{sec:Completeness}).

A main motivation for the development of the conflict resolution calculus was that it might eventually serve as a theoretical common ground for existing first-order provers that try to harness or mimic the power of SAT-solvers 
%(e.g. \cite{BonacinaPlaisted,Weidenbach,Baumgartner,BaumgartnerTinelli,AVATAR} 
(cf. Section \ref{sec:RelatedWork})) or as a starting point for the development of new provers, in the same way that the pure resolution calculus provided the basic foundation for several generations of automated theorem provers in the last decades. To achieve this goal, the calculus is presented in a general way, avoiding premature optimizations and refinements, so that future work may easily build on it and explore various proof search strategies and implementation techniques. 
%We critically assess (in Section \ref{sec:Engineering}) some issues that might make the implementation of \UPRCL and of proof search strategies based on it harder and less efficient than in the propositional case.

\section{Recalling Resolution}
\label{sec:Resolution}

\emph{Clauses} (denoted $c$, possibly subscripted) are disjunctions of literals. A \emph{literal} is either an atom or a negated atom, and an \emph{atom} is a $n$-ary predicate (denoted $P$ or $Q$) applied to $n$ terms. A \emph{term} is either a constant (denoted $a$ or $b$), a variable (denoted $x$, $y$, $v$ or $z$) or an $n$-ary function (denoted $f$ or $g$) applied to $n$ terms. Variables in a clause are assumed to be implicitly universally quantified. A clause having a single literal is called \emph{unit}. If $\ell$ is a literal, $\dual{\ell}$ denotes its dual (i.e. $\dual{P} = \neg P$ and $\dual{\neg P} = P$). The nullary atoms $\top$ (\emph{verum}) and $\bot$ (\emph{falsum}) have special meanings characterized by the following equations: $\Gamma \vee \bot = \Gamma$ and $\Gamma \vee \top = \top$. All inference rules operating on clauses are assumed to be modulo disjunction's associativity and commutativity, modulo negation's involutivity and modulo the equations for $\top$ and $\bot$. The empty clause is logically equivalent to the clause containing only $\bot$. Therefore, slightly abusing notation, it is denoted by $\bot$. Substitutions (denoted by $\sigma$, possibly sub- and superscripted) are assumed to implicitly avoid variable capture. The empty (i.e. identity) substitution is denoted $\varepsilon$.

The inference rules of the resolution calculus are shown in Fig. \ref{fig:Resolution}. A resolution \emph{proof} of a clause $c$ from a set of clauses $S$ is a directed acyclic graph (DAG) where leaves (i.e. input nodes) are clauses from $S$, internal nodes are obtained from their parents through application of the inference rules and the sink node is the clause $c$. 
A resolution \emph{refutation} of a set of clauses $S$ is a proof of the empty clause (denoted $\bot$) from $S$.
It is assumed that distinct input clauses do not share variables. Furthermore, the inference rules implicitly generate fresh symbols for variables, thereby maintaining the invariant that distinct clauses do not share variables.

Proof DAGs are sometimes displayed as a collection of trees according to the following convention: nodes used as premises more than once are given names (e.g. $\varphi$, $\psi$ or $\xi$) when they are used for the first time, and the names are used to refer to the nodes whenever they are used again. By naming and referring, wide proof trees can also be broken down in smaller displayable parts.

\begin{example}
Consider a proof with the following non-tree form: 

\begin{center}
\includegraphics[width=0.15\textwidth]{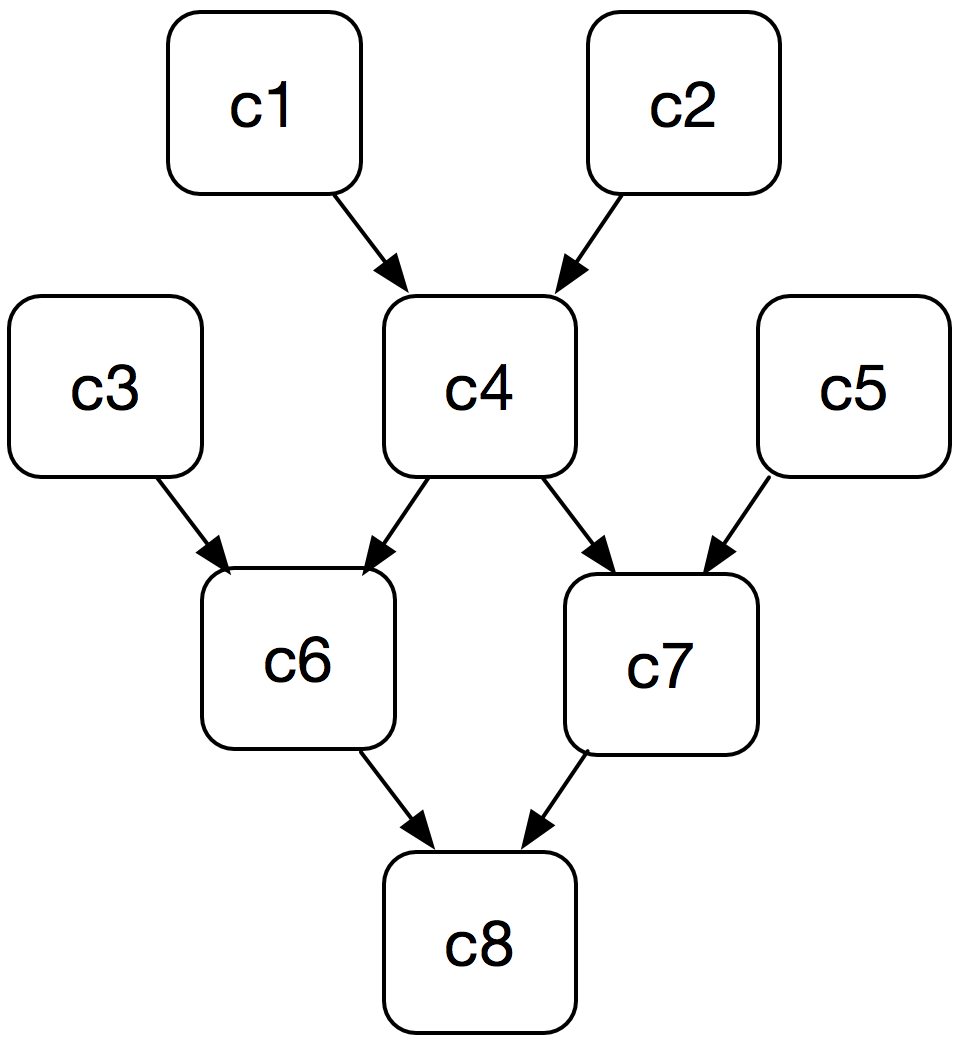}
\end{center}

\noindent
It can be displayed as the single tree with names and references below, where the second (rightmost) occurrence of the name $\psi$ is to be understood as a reference to the node named $\psi$ by the first (leftmost) occurrence of $\psi$:
\begin{prooftree}
\AXC{$c_3$}
		\AXC{$c_1$}
				\AXC{$c_2$}
			\BIC{$\psi: c_4$}
	\BIC{$c_6$}
						\AXC{$\psi$}
								\AXC{$c_5$}
							\BIC{$c_7$}
				\BIC{$c_8$}
\end{prooftree}	
Or it can also be displayed as the following forest, where the two occurrences of the name $\psi$ in the lower tree are to be understood as references to the node named $\psi$ in the upper tree:
\begin{prooftree}
		\AXC{$c_1$}
				\AXC{$c_2$}
			\BIC{$\psi: c_4$}
\end{prooftree}	
\begin{prooftree}
\AXC{$c_3$}
			\AXC{$\psi$}
	\BIC{$c_6$}
						\AXC{$\psi$}
								\AXC{$c_5$}
							\BIC{$c_7$}
				\BIC{$c_8$}
\end{prooftree}	
\end{example}

Given a set of clauses, a resolution prover exhaustively applies the inference rules, generating more and more clauses. If the initial clause set is unsatisfiable and a fair clause/rule selection strategy is used, the empty clause is eventually derived, because resolution is refutationally complete \cite{Robinson}. If the set is satisfiable, the prover will either never terminate or will terminate in a state where the set of initial and derived clauses is saturated with respect to redundancy criteria (i.e. only redundant clauses would still be derivable) (cf. \citeauthor{WaldmannEPSSaturation} \citeyear{WaldmannEPSSaturation}). 

One practical problem in this saturation approach is the vast number of clauses that are generated. This led to research on refinements of the resolution calculus, aiming at restricting the inference rules in order to generate fewer clauses, and on efficient ways to detect and delete redundant (e.g. subsumed) clauses. These efforts culminated in the \emph{superposition}\footnote{\UPRCL is based on resolution instead of superposition, because superposition's ordering-based refinements would restrict unit-propagation and the selection of decision literals. In SAT-solvers unit-propagation is unrestricted (because it is very efficient anyway) and the best literal selection strategies are not based on orderings. By extending unrestricted resolution, \UPRCL remains general enough to admit the strategies used by SAT-solvers.} calculus \cite{BachmairGanzinger1990,BachmairGanzinger1994,WaldmannEPSSuperposition}, which extends the resolution calculus with a paramodulation rule \cite{Paramodulation} for equality reasoning and refines it with ordering restrictions on terms and literals.

\newcommand{\res}[1]{\mathbf{r}(#1)}
\newcommand{\factoring}[1]{\mathbf{f}(#1)}

\begin{figure}
\begin{calculus}
\centering

\textbf{Resolution:}
$$
\infer[\res{\sigma}]{(\Gamma \vee \Delta)~\sigma}{\Gamma \vee \ell & \dual{\ell'} \vee \Delta}
$$

where $\sigma$ is a unifier of $\ell$ and $\ell'$.

\bigskip

\textbf{Factoring:}
$$
\infer[\factoring{\sigma}]{(\ell \vee \ell'_1 \vee \ldots \vee \ell'_m)~\sigma}{\ell_1 \vee \ldots \vee \ell_n \vee \ell'_1 \vee \ldots \vee \ell'_m}
$$

where $\sigma$ is a unifier of $\ell_1$, \ldots $\ell_n$ and $\ell = \ell_k~\sigma$, for any $k \in \{1, \ldots, n \}$.

\end{calculus}
\caption{Resolution Calculus}
\label{fig:Resolution}
\end{figure}

Another practical problem is that the resolvent of a clause with $n$ literals and another clause with $m$ literals has $n + m - 2$ literals. When iterated, this results in very long clauses and, consequently, a loss of efficiency. This practical problem has been solved with a technique known as \emph{splitting} \cite{WeidenbachSplitting}: if the current set of clauses is $S \cup \{ \Gamma_1 \vee \ldots \vee \Gamma_k \}$ and the sets of variables $V_i$ of $\Gamma_i$ are mutually disjoint, then we can split the long clause $\Gamma_1 \vee \ldots \vee \Gamma_k$ into its variable-disjoint components and the clause set into the $k$ sets $S \cup \{\Gamma_i\}$ (for $1 \leq i \leq k$). The disjointness of the sets of variables $V_i$ ensures that we can check the unsatisfiability of each resulting clause set independently: $S \cup \{ \Gamma_1 \vee \ldots \vee \Gamma_k \}$ is unsatisfiable iff $S \cup \{\Gamma_i\}$ is unsatisfiable for every variable-disjoint component $\Gamma_i$.

From a proof-theoretical perspective, splitting resembles the $\beta$-rule of free-variable tableaux \cite{Tableaux,SPASSManual}. Therefore, superposition provers that implement splitting \cite{SPASS,E,Vampire} can be seen as hybrids combining resolution/superposition and tableaux. Up to now, however, there has been no single pure proof system capable of characterizing what is going on inside a modern state-of-the-art first-order theorem prover. This gap between theory and practice is something that can be remedied with the adoption of the \UPRCL calculus proposed here (cf. Section \ref{sec:Splitting}).

\section{Recalling DPLL and CDCL}
\label{sec:DPLLCDCL}

In the propositional case, Davis, Logemann and Loveland \citeyear{DLL} had already noticed that the propositional resolution rule \cite{DavisPutnam} ``can easily increase the number and the lengths of the clauses'' and proposed to replace it by a form of splitting, which is, however, different from the later notion of splitting described in Section \ref{sec:Resolution}. Instead of splitting a clause into variable-disjoint components, we select a propositional atom $P$ and split the problem in two subproblems: one where $P$ is assumed to be true and the other where it is assumed to be false. Nowadays, the so-called DPLL procedure is presented slightly differently, but equivalently. We \emph{decide} to assign the truth value \texttt{true} (or \texttt{false}) to an atom; then, through \emph{unit propagation}, other atoms will be assigned truth values as well. Repeating this process of decisions and propagations, we will either reach an assignment that satisfies all clauses (if the clause set is satisfiable) or we will reach a \emph{conflict} where we are to assign both \texttt{true} and \texttt{false} to an atom. In the latter case, we backtrack some of our decisions, and try different assignments.

In contrast to saturation-based theorem proving, DPLL-based sat-solving does not generate any clause at all. But this is, of course, dependent on the fact that in propositional logic it suffices to consider only two truth-value assignments for each atom. In a na\"ive adaptation of this idea to first-order logic, on the other hand, we would need to consider truth-value assignments for each instance of an atom containing variables. We would need to generate possibly several\footnote{By Herbrand's theorem, a finite number of instances would suffice in the case of an unsatisfiable clause set.} instances.

In practice, it has been found that it is, nevertheless, beneficial to generate \emph{some} clauses when backtracking from conflicts. For example, suppose that the backtracking DPPL procedure decided to assign \texttt{true} to $P$ and $Q$, and this led to a conflict. It is then forced to backtrack these decisions and try other decisions. Without clause learning, it could happen that, after assigning truth values to other atoms, it would again consider the possibility of assigning \texttt{true} to $P$ and $Q$, even though it is clear (from the previous conflict) that $P$ and $Q$ cannot be both \texttt{true}, independently of later assignments to other atoms. To prevent this from happening, we can generate and add the clause $\neg P \vee \neg Q$ to the set of clauses. Then, whenever the procedure retries assigning, for instance, \texttt{true} to $P$ it will immediately conclude (by unit propagation) that \texttt{false} should be assigned to $Q$. This idea is known as \emph{conflict-driven clause learning}.

The procedure up to a conflict can be understood as the construction of a directed graph. Nodes are literals which have been assigned \texttt{true}. A \emph{decision literal} (i.e. a literal with truth value assigned by decision) has no incoming edge. A \emph{propagated literal} (i.e. a literal with truth value assigned by unit propagation) $\ell$ has incoming edges $(\ell_i,\ell)$ for $0 < i \leq n$ iff the clause $\dual{\ell_1} \vee \ldots \vee \dual{\ell_n} \vee \ell$ was the clause used by unit propagation to assign a truth value to $\ell$. A conflict is indicated by the simultaneous presence of any literal and its dual in the graph. When a conflict is detected, the graph can be analyzed to determine clauses that should be learned. Various conflict analysis algorithms exist \cite{GRASP,CDCL}. The conceptually simplest one recommends learning a clause that is a disjunction of the negations of the decision literals. More sophisticated algorithms \cite{FirstUIP} are capable of learning stronger clauses. An important benefit of conflict-driven clause learning is that redundant (i.e. subsumed) clauses are never derived.

The learned clause can be derived by a sequence of resolution steps using the clauses corresponding to the edges in the graph as premises. When this is done, a SAT-solver is capable of outputting a propositional resolution refutation for an unsatisfiable clause set \cite{BiereTraceCheck}. However, most developers of SAT-solvers consider the overhead (in both proving time and memory consumption) of doing so unacceptable, especially when advanced techniques for minimizing learned clauses are used. Instead, they prefer to generate proof certificates in the DRUP or DRAT formats \cite{DRATTrim}, which record clauses that have been learned, but do not inform which premises are needed to derive them. A consequence of this lack of information is that checking a DRUP/DRAT certificate or converting it to a resolution refutation (using the DRAT-Trim tool) can take as long as solving the problem in the first place.

\begin{example}
\label{example:ConflictGraphs}
Consider the clause set $\{ P \vee Q , \ P \vee \neg Q , \ \neg P \vee Q , \ \neg P \vee \neg Q \}$. Deciding $P$ and propagating units results in the conflict graph at the left side below. We backtrack and learn the unit clause $\neg P$, whose propagation leads to the conflict graph in the right side below. Since this last conflict does not depend on any decision literal, no backtracking is possible, and we may conclude that the clause set is unsatisfiable. 

\smallskip

\noindent
\includegraphics[width=0.48\textwidth]{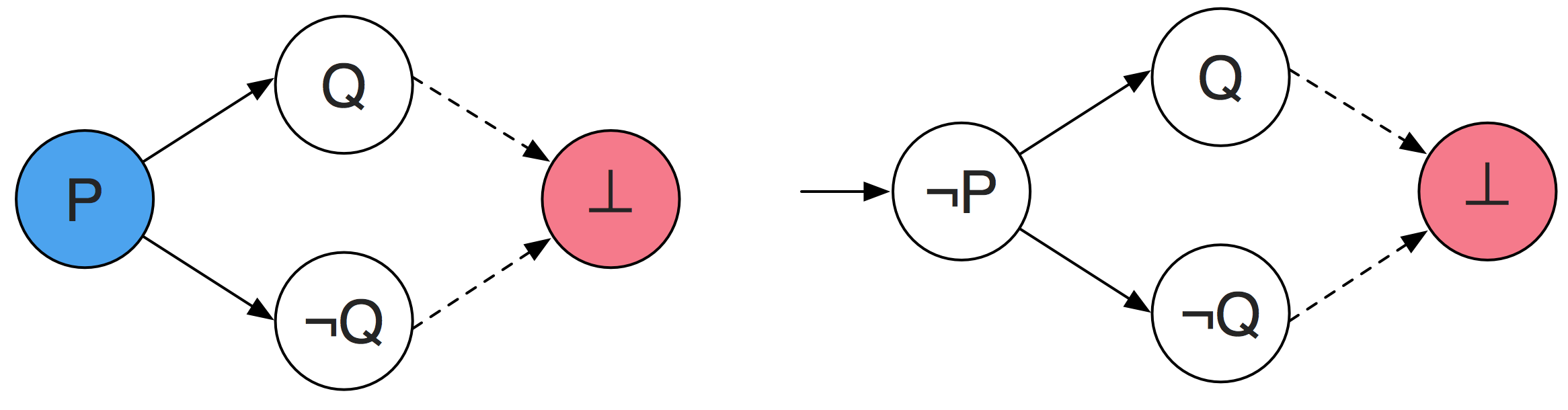}

\smallskip

\noindent
The resolution proof extracted from the first conflict graph is:
\begin{footnotesize}
\begin{prooftree}
\AXC{$\neg P \vee Q$}
		\AXC{$\neg P \vee \neg Q$}
	\BIC{$\neg P \vee \neg P$}
	\UIC{$\neg P$}
\end{prooftree}
\end{footnotesize}
The resolution proof extracted from the second conflict graph is:
\begin{footnotesize}
\begin{prooftree}
\AXC{$P \vee Q$}
		\AXC{$P \vee \neg Q$}
	\BIC{$P \vee P$}
	\UIC{$P$}
			\AXC{$\neg P$}
		\BIC{$\bot$}
\end{prooftree}
\end{footnotesize}
\end{example}

\section{The Conflict Resolution Calculus}
\label{sec:UPRCL}

\newcommand{\upr}[1]{\mathbf{u}(#1)}
\newcommand{\con}[1]{\mathbf{c}(#1)}
\newcommand{\cdcl}{\mathbf{cl}}

As we have seen in the previous two sections, both propositional and first-order automated deduction have progressed (in different ways) much beyond their historical common roots in resolution. Techniques such as splitting, conflict graphs and conflict-driven clause learning are not so easily explained in terms of a pure resolution calculus. There is a growing gap between the current state-of-the-art in automated deduction and its original proof-theoretical foundation. In this section, we propose the \UPRCL calculus, which modifies the first-order resolution calculus by incorporating ideas from SAT-solving, in an attempt to reduce not only the gap between automated deduction and proof theory but also between the first-order and the propositional cases. 

\begin{figure}
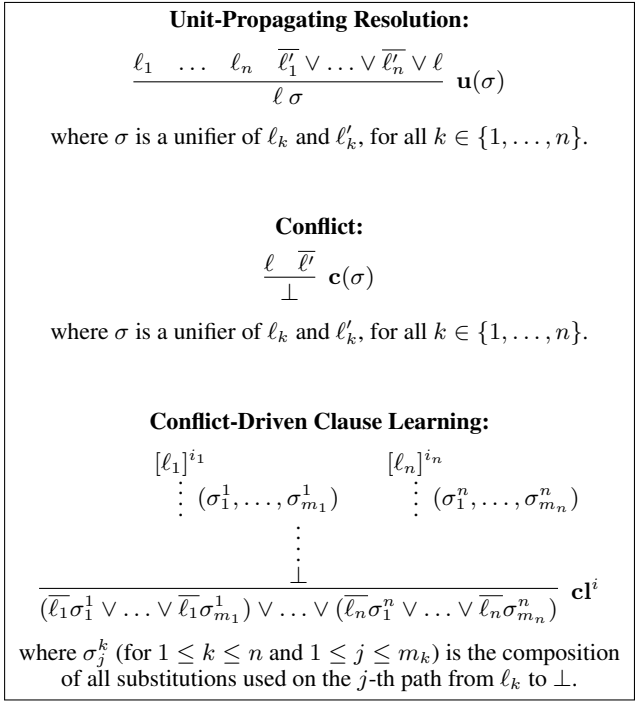

\begin{calculus}
\centering

\textbf{Unit-Propagating Resolution:}
$$
\infer[\upr{\sigma}]{\ell~\sigma}{\ell_1 & \ldots & \ell_n & \dual{\ell'_1} \vee \ldots \vee \dual{\ell'_n} \vee \ell}
$$

where $\sigma$ is a unifier of $\ell_k$ and $\ell'_k$, for all $k \in \{1, \ldots, n \}$.

\bigskip
\bigskip

\textbf{Conflict:}
$$
\infer[\con{\sigma}]{\bot}{\ell & \dual{\ell'}}
$$

where $\sigma$ is a unifier of $\ell_k$ and $\ell'_k$, for all $k \in \{1, \ldots, n \}$.

\bigskip
\bigskip

\textbf{Conflict-Driven Clause Learning:}
$$
\infer[\cdcl^{i}]
{(\dual{\ell_1} \sigma^1_1 \vee \ldots \vee \dual{\ell_1} \sigma^1_{m_1}) \vee \ldots \vee (\dual{\ell_n} \sigma^n_1 \vee \ldots \vee \dual{\ell_n} \sigma^n_{m_n})}
{\infer*{\bot}{\infer*[(\sigma_1^1,\ldots,\sigma_{m_1}^1)]{}{[\ell_1]^{i_1} } &  & \infer*[(\sigma_1^n,\ldots,\sigma_{m_n}^n)]{}{[\ell_n]^{i_n} } }}
$$

where $\sigma^k_j$ (for $1 \leq k \leq n$ and $1 \leq j \leq m_k$) is the composition of all substitutions used on the $j$-th path from $\ell_k$ to $\bot$.

\end{calculus}
\caption{The Conflict Resolution Calculus \UPRCL}
\label{fig:UPRCL}
\end{figure}

As in resolution, a \UPRCL \emph{derivation} is a directed acyclic graph where nodes are clauses and internal nodes are obtained from their parents by one of the inference rules shown in Fig.~\ref{fig:UPRCL}. The \emph{conflict} rule is just a restriction of the resolution rule. The \emph{unit-propagating resolution}\footnote{This rule is also known as \emph{unit-resulting resolution} \cite{UnitResultingResolution,Prover9Manual}. Here we use the name \emph{unit-propagating resolution} instead in order to make the connection with the technique of unit-propagation more explicit.} rule is essentially a sequence of applications of the resolution rule where the left premises must always be unit clauses; the conclusion clause must be unit as well, and its literal is called a \emph{propagated literal}.

The main innovation lies in the \emph{conflict-driven clause learning} rule. The literals within brackets are the \emph{decision literals} that have been assumed. The superscript index $i$ indicates that this assumption is discharged by the $\cdcl$ inference with index $i$. 
It is not required that a $\cdcl$ inference discharge all decision literals above it. Some decision literals may be left undischarged, to be discharged by future $\cdcl$ inferences. The vertical dots denote any derivation of $\bot$ using the decision literals, input clauses and previously derived clauses. The conclusion clause of this rule is the \emph{learned clause}. In contrast to the propositional case, the learned clause must be a disjunction of negations of \emph{instances} of the discharged decision literals, because variables occurring in the discharged decision literals may be instantiated by unifications performed during the proof. Since the derivation of $\bot$ need not be tree-like, we may need to consider several instances of each decision literal.

%ToDo: the following idea would be an interesting generalization of UPRCL, but it would complicate the proof of soundness. Therefore, I leave it out for the moment.
%An assumption may be discharged by more than one $\cdcl$ inference, in which case it will a superscript index for each $\cdcl$ inference that discharges it. 

A \UPRCL derivation is a \UPRCL \emph{proof} iff all its decision literals have been discharged. A \UPRCL \emph{refutation} is a \UPRCL proof of $\bot$.

Resolution's \emph{factoring} rule can be simulated by a sequence of decisions, one unit-propagation, one conflict and one conflict-driven clause learning. In this way, we can prove the following lemma.

\begin{lemma}
Resolution's factoring rule is admissible in \UPRCL.
\end{lemma}
\begin{proof}
Let $\varphi'$ be a \UPRCL derivation of $\ell_1 \vee \ldots \vee \ell_n \vee \ell'_1 \vee \ldots \vee \ell'_m$ and consider constructing $\varphi$ by applying the factoring inference to the conclusion of $\varphi'$, as shown below:
$$
\infer[\factoring{\sigma}]{(\ell \vee \ell'_1 \vee \ldots \vee \ell'_m)~\sigma}{\infer*[\varphi']{\ell_1 \vee \ldots \vee \ell_n \vee \ell'_1 \vee \ldots \vee \ell'_m}{} }  
$$
This is admissible because, instead of using the factoring inference, we could have used a sequence of \UPRCL inferences, as shown in Fig.~\ref{fig:FactoringAdmissibility}.

\begin{figure*}
\centering
$$
\infer[\cdcl^1]
{(\ell \vee \ell'_1 \vee \ldots \vee \ell'_m)~\sigma}
{
	\infer[\con{\varepsilon}]{\bot}
	{
		\infer[\upr{\sigma}]{\ell'_m~\sigma}
		{
			\psi: [\dual{\ell}]^{1_1}
			&
			\overbrace{\psi \quad \ldots \quad \psi}^{n-1 \ \mathrm{times}}
			& 
			[\dual{\ell'_1}]^{1_{n+1}}
			&
			\ldots
			&
			[\dual{\ell'_{m-1} }]^{1_{n+m-1}}
			&
			\infer*[\varphi']{\ell_1 \vee \ldots \vee \ell_n \vee \ell'_1 \vee \ldots \vee \ell'_m}{}
		}
		&
		[\dual{\ell'_m}~\sigma]
	}
}
$$

\smallskip

where $\sigma$ is a unifier of $\ell_1$, \ldots $\ell_n$ and $\ell = \ell_k~\sigma$, for any $k \in \{1, \ldots, n \}$.
\caption{Simulation of a Factoring Inference in \UPRCL}
\label{fig:FactoringAdmissibility}
\end{figure*}

\end{proof}

The simulation of factoring depends on a sufficient degree of freedom in the choice of decision literals. We must be allowed (as indeed we are in \UPRCL) to assume a decision literal ($\dual{\ell}$) that is the dual of an instance of all $\ell_i$ (for $1 \leq i \leq n$). 

% The following paragraph is wrong. This is something that might be interesting to explore in more detail in an extended version of this paper.
%The simulation also depends on the fact that \UPRCL derivations may be directed DAGs that are not tree-like. We use the decision literal $\dual{\ell}$ $n$ times in the unit-propagating resolution inference, but we assume it only once. And because we assume it only once, its dual ($\ell$) appears only once in the conclusion of the $\cdcl$ inference.

\newcommand{\sub}[2]{#1 \backslash #2}

\begin{example}
\label{example:UPRCL}
Consider the following clause set:
$$ \{ P(z) \vee Q , \ P(y) \vee \neg Q , \ \neg P(a) \vee Q , \ \neg P(b) \vee \neg Q \}$$

\noindent
It admits the \UPRCL refutation shown in Fig. \ref{fig:UPRCLRefutation}. A shorter refutation would be possible if we had taken, for instance, $Q$ as a decision literal. But taking $P(x)$ as a decision literal instead, as done in the refutation in Fig. \ref{fig:UPRCLRefutation}, we can see how conflict driven clause learning behaves in the first-order case, when decision literals can contain variables, that can be instantiated during the process of propagation. In one path from $\psi^1$ to $\bot$ just above the $\cdcl^1$ inference, the unification performed by the unit-propagating resolution inference instantiates $x$ with $a$, whereas in the other path $x$ is instantiated with $b$. Therefore, the $\cdcl^1$ inference learns the clause $\neg P(a) \vee \neg P(b)$, which is the disjunction of the negations of all the instances of the decision literal $P(x)$. This is in contrast with (and a generalization of) the propositional case, where instances did not need to be considered.

\begin{figure*}
%\begin{tiny}
\begin{prooftree}
\AXC{$\psi_11 : [P(x)]^1$}
		\AXC{$\xi_1: \neg P(a) \vee Q$} \RightLabel{$\upr{\{ \sub{x}{a} \} }$}
	\BIC{$Q$}
			\AXC{$\psi_1$}
					\AXC{$\neg P(b) \vee \neg Q$} \RightLabel{$\upr{\{ \sub{x}{b} \} }$}
				\BIC{$\neg Q$} \RightLabel{$\con{\varepsilon}$}
		\BIC{$\bot$} \RightLabel{$\cdcl^1$}
		\UIC{$\varphi_1: \neg P(a) \vee \neg P(b)$}
\end{prooftree}

\begin{prooftree}
						\AXC{$\psi_2: [\neg P(a)]^2$}
								\AXC{$ P(z) \vee Q $} \RightLabel{$\upr{ \{ \sub{z}{a} \} }$}
							\BIC{$Q$}
									\AXC{$\psi_2$}
											\AXC{$ \xi_2: P(y) \vee \neg Q$} \RightLabel{$\upr{ \{ \sub{y}{a} \} }$}
										\BIC{$\neg Q$} \RightLabel{$\con{\varepsilon}$}
								\BIC{$\bot$} \RightLabel{$\cdcl^2$}
								\UIC{$\varphi_2: P(a)$}	  
\end{prooftree}

\begin{prooftree}
		\AXC{$\varphi_2$}	
					\AXC{$\varphi_1$} \RightLabel{$\upr{\varepsilon}$}				            		
				\BIC{$\neg P(b)$} 
					    \AXC{$\xi_2$} \RightLabel{$\upr{\{ \sub{y}{b} \} }$}
					\BIC{$\neg Q$}
							\AXC{$\varphi_2$}
									\AXC{$\xi_1$} \RightLabel{$\upr{\varepsilon}$}	
								\BIC{$Q$} \RightLabel{$\con{\sigma}$}
						\BIC{$\bot$}
\end{prooftree}
%\end{tiny}
\caption{\UPRCL Refutation for the clause set from Example \ref{example:UPRCL}.}
\label{fig:UPRCLRefutation}
\end{figure*}

As in the propositional case, the decisions and unit propagations can be represented graphically:
\medskip

\noindent
\includegraphics[width=0.48\textwidth]{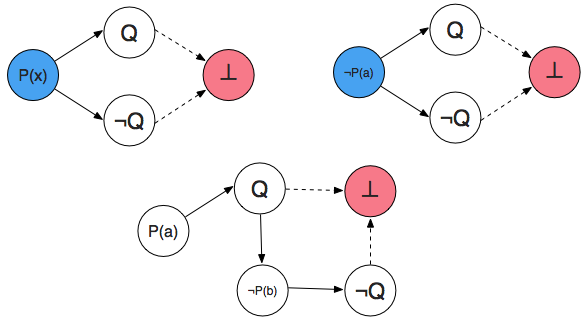}
\end{example}

\subsection{An Isomorphism between Conflict Graphs and Single-Conflict Sub-Derivations in Conflict Resolution}
\label{sec:Isomorphism}

By comparing the conflict graphs and \UPRCL derivations in Example \ref{example:UPRCL}, it is noticeable that there is a straightforward isomorphism between conflict graphs and \UPRCL sub-derivations with a single conflict inference. Every decision literal in a conflict graph appears as a decision literal in the corresponding \UPRCL derivation. Every propagated literal in the conflict graph appears as a propagated literal derived by a unit-propagating resolution inference, and the clause associated to the incoming edges of the propagated literal is exactly the non-unit clause used as the rightmost premise of the unit-propagating inference. Finally, the conflict in the conflict graph is a conflict inference in the corresponding \UPRCL derivation. 

In contrast, the correspondence between resolution derivations and conflict graphs is imperfect. As illustrated in Example \ref{example:ConflictGraphs}, we have a map from conflict graphs to resolution derivations; however, this map is not an isomorphism, simply because it is not even surjective. Furthermore, there is a mismatch between the conflict graph operations (i.e. decisions, propagations and conflict) and the operations of the resolution calculus (i.e. the resolution and factoring inference rules). In other words, no map from conflict graphs to resolution derivations could be an isomorphism, because the algebraic structure cannot be preserved. From this algebraic point of view, we may conjecture that the popular belief that (propositional) resolution is the underlying proof system of modern SAT-solvers (which actually implement the CDCL procedure based on conflict graphs) is mistaken. We also speculate that the mismatch is the theoretical explanation for the overhead experienced in the transformation of conflict graphs to resolution derivations (as discussed in the end of Section \ref{sec:DPLLCDCL}). Perhaps a calculus such as \UPRCL, that enjoys a better correspondence to conflict graphs, could enable proof production with less overhead.

\section{Refutational Completeness}
\label{sec:Completeness}

A proof system \textbf{P} is \emph{refutationally complete} iff any unsatisfiable clause set has a refutation in \textbf{P}. Instead of proving refutational completeness for \UPRCL directly, we will prove it indirectly, showing that \UPRCL can simulate another refutationally complete proof system. A proof system \textbf{P} simulates another proof system \textbf{Q} iff there is a map transforming any \textbf{Q}-derivation of $c$ from $S$ to \textbf{P}-derivation of $c$ from $S$.

This indirect approach to proving completeness can be traced back at least to Gentzen's work (\citeyear{Gentzen}), who applied it to his natural deduction and sequent calculi. In our case, the target proof system for the simulation is resolution, and the key idea of the simulation is that every resolution step that is not a unit-propagating resolution inference can be simulated by several decisions, two unit-propagating resolution inferences, one conflict inference and one conflict-driven clause learning inference.

\begin{theorem}
\label{theorem:SimulationOfResolution}
\UPRCL linearly simulates Resolution.
\end{theorem}
\begin{proof} 
Let $\psi$ be a Resolution derivation of a clause $c$ from a set of clauses $S$. We show that there is a \UPRCL derivation $\varphi$ of $c$ from $S$, proceeding by induction:
\begin{itemize}
\item \emph{Base Case:} $\psi$ is just a single node $c$. In this case, $\varphi$ is just the single node $c$ as well.
\item \emph{Induction Case 1:} $\psi$ ends with a factoring inference $\rho$. In this case, let $\psi'$ be the subderivation whose conclusion $c'$ is the premise of $\rho$. By induction hypothesis, there is a \UPRCL derivation $\varphi'$ of $c'$ from $S$. And then $\varphi$ can be constructed as the \UPRCL derivation of $c$ from $S$ obtained from $\varphi'$ by applying the admissible factoring inference rule to its conclusion in the same way as $\rho$ in $\psi$ or by simulating factoring as shown in Fig.~\ref{fig:FactoringAdmissibility}. In any case, the conclusion of $\varphi$ is $c$, as desired.
\item \emph{Induction Case 2:} $\psi$ ends with a resolution inference. In this case, $\psi$ is of the following form:
$$
\infer[\res{\sigma}]
{(\ell_1 \vee \ldots \vee \ell_n \vee \ell'_1 \vee \ldots \vee \ell'_m)~\sigma}
{\infer*[\psi_1]{\ell_1 \vee \ldots \vee \ell_n \vee \ell \qquad}{} & 
 \infer*[\psi_2]{\qquad \dual{\ell'} \vee \ell'_1 \vee \ldots \vee \ell'_m}{} }
$$
By induction hypothesis, we have a \UPRCL derivation $\varphi_1$ of $\ell_1 \vee \ldots \vee \ell_n \vee \ell$ from $S$ and a \UPRCL derivation $\varphi_2$ of $\dual{\ell'} \vee \ell'_1 \vee \ldots \vee \ell'_m$ from $S$. Then a \UPRCL derivation $\varphi$ of $(\ell_1 \vee \ldots \vee \ell_n \vee \ell'_1 \vee \ldots \vee \ell'_m)~\sigma$ can be constructed as shown in Fig.~\ref{fig:SimulationOfResolution}.

\begin{figure*}
$$
\infer[\cdcl^1]{(\ell_1 \vee \ldots \vee \ell_n \vee \ell'_1 \vee \ldots \vee \ell'_m)~\sigma}{
	\infer[\con{\sigma}]{\bot}{
		\infer[\upr{\varepsilon}]{\ell}{[\dual{\ell_1}]^1 & \ldots & [\dual{\ell_n}]^1 & \infer*[\varphi_1]{\ell_1 \vee \ldots \vee \ell_n \vee \ell}{} }
		&
		\infer[\upr{\varepsilon}]{\dual{\ell'}}{ [\ell'_1]^1 & \ldots & [\ell'_m]^1 & \infer*[\varphi_2]{\dual{\ell'} \vee \ell'_1 \vee \ldots \vee \ell'_m}{}}
	}
}
$$
\caption{Simulation of a Resolution Inference in \UPRCL}
\label{fig:SimulationOfResolution}
\end{figure*}
\end{itemize}

The simulation is linear both in length (i.e. number of inferences) and size (i.e. number of literals). If $\psi$ has $n$ resolutions and $m$ factorings, then $\varphi$ has $n$ clause learning inferences, $n$ conflicts, $2$ unit propagations and $m$ factorings. Hence, $\mathit{length}(\varphi) = 2n+m+2 \in O(n+m) = O(\mathit{length}(\varphi))$. If $\psi$ has $n'$ literals occurring in conclusions of resolution inferences and $m'$ literals occurring in conclusions of factoring inferences, then $\varphi$ has $n'$ logical symbols occurring in conclusions of clause learning inferences, $2 n$ logical symbols occurring in premises of conflict inferences, $n'$ literals occurring as decision literals for unit propagations and $m'$ literals. Hence, $\mathit{size}(\varphi) = 2n' + 3n + m'$. Since every resolution inference in $\psi$ has at least one literal in its conclusion, except for the last one deriving the empty clause, $3n \leq 3n' + 1$. Therefore, $\mathit{size}(\varphi) \in O(5n' + 1 + m')$ and thus $\mathit{size}(\varphi) \in O(\mathit{size}(\psi))$.
\end{proof}

\begin{corollary}
\UPRCL is refutationally complete.
\end{corollary}
\begin{proof}
Let $C$ be an unsatisfiable clause set. As resolution is a refutationally complete calculus \cite{Robinson}, there is a resolution refutation $\psi$ of $C$. By Theorem \ref{theorem:SimulationOfResolution}, $\psi$ can be transformed to a \UPRCL refutation $\varphi$ of $C$.
\end{proof}

A mere restriction of resolution to unit-propagating resolution would result in a refutationally incomplete calculus. The unsatisfiable clause sets from Examples \ref{example:ConflictGraphs} and \ref{example:UPRCL}, for instance, would not be refutable. By incorporating decision literals, as well as the conflict rule and the conflict-driven clause learning rule, we regain refutational completeness.

The fact that we need two unit-propagating resolution inferences, one conflict and one conflict-driven clause learning to simulate a single resolution inference (as shown in Fig.~\ref{fig:SimulationOfResolution}) may lead us to think that \UPRCL is more bureaucratic and more inefficient than resolution. However, efficiency of proof search is not directly correlated with proof length. The efficiency of \UPRCL is a consequence of the fact that much fewer clauses are generated by unit-propagating resolution than by unrestricted resolution and the clause sizes are reduced through decisions and propagations. Moreover, in any case, any resolution proof search, as well as any resolution proof, can be simulated in the \UPRCL calculus with only a (small) linear increase in length.

%ToDo: do we retain completeness if we restrict decision literals only to literals occurring in the input clauses?

\section{Soundness}
\label{sec:Soundness}

\newcommand{\ND}{\textbf{ND}\xspace}
\newcommand{\CND}{\textbf{CND}\xspace}

To prove soundness, we exploit the key observation that decision literals resemble natural deduction's \emph{assumptions} and conflict-driven clause learning resembles \emph{implication/negation introduction}. Therefore, natural deduction is an excellent candidate for proving soundness indirectly, again by simulation. However, typical natural deduction rules operate on general formulas, which are not necessarily in clause form, and this makes a direct simulation technically difficult. In order to overcome this challenge, we define an intermediary \emph{clausal natural deduction calculus} (abbreviated as \CND) with inference rules that operate on clauses, as shown in Fig.~\ref{fig:CND}.

\begin{figure}
\begin{calculus}
\centering

\textbf{Implication Elimination (Modus Ponens):}
$$
\infer[\imp_E]{\Gamma}{\ell & \dual{\ell} \vee \Gamma}
$$

\medskip

\textbf{Implication Introduction:}
$$
\infer[\imp_I^i]
{\dual{\ell} \vee \Gamma}
{\infer*{\Gamma}{[\ell]^i}}
$$

\medskip

\textbf{Universal Quantification Elimination:}
$$
\infer[\forall_E]{\Gamma \sigma }{\Gamma}
$$

\medskip

\textbf{Universal Quantification Introduction:}
$$
\infer[\forall_I]
{\Gamma}
{\Gamma \{ \sub{x_1}{\alpha_1}, \ldots, \sub{x_n}{\alpha_n}\} }
$$

$\alpha_k$ must be a distinct \emph{eigen-variable}: \\ it should occur neither in $\Gamma$ nor in any undischarged assumption.

\end{calculus}
\caption{The Clausal Natural Deduction Calculus \CND}
\label{fig:CND}
\end{figure}

The clausal natural deduction calculus \CND can be simulated by any standard non-clausal natural deduction calculus extended with a classical rule for double negation elimination\footnote{An example of such a natural deduction calculus is shown in the appendix}. The key idea is to use the well-known classical equivalence $A \vee B \equiv (\dot{\neg} A \imp B)$ (where $\dot{\neg} A$ is an abbreviation for $A \imp \bot$), in order to transform the clauses in a $\CND$ proof into formulas containing only implication, which are therefore suitable for a minimal non-clausal natural deduction calculus. When transforming a \CND proof into a standard non-clausal natural deduction proof, sequences of implication introduction/elimination rules may have to be added to the natural deduction proof, in order to reorder literals (because associativity and commutativity of disjunction is implicitly taken into account by \CND's inference rules, but must be handled explicitly in a standard natural deduction calculus). The classical rule of double negation elimination is needed in order to handle the involutivity of classical negation, which is implicit in \CND\footnote{\CND is a calculus for \emph{classical} logic: the law of excluded middle can be easily derived with a single application of implication introduction. Interestingly, its classicality is implicit in the use of involutive negation and the equivalence involving disjunctions and implications.}. A more detailed proof of this simulation is omitted because it would be tedious and space-consuming. Soundness of \CND is a corollary of the simulation, since natural deduction is sound.

Remembering that all clausal rules are assumed to be modulo negation's involutivity and modulo the neutrality of $\bot$ w.r.t. disjunction, the rules for negation introduction and elimination shown in Fig.~\ref{fig:CNDNegationRules} are admissible in \CND, since they are just special cases of, respectively, implication introduction and elimination, when $\Gamma = \bot$. We are now ready to prove the following theorem.

\begin{figure}
\begin{calculus}
\centering
\textbf{Negation Elimination:}
$$
\infer[\neg_E]{\bot}{\ell & \dual{\ell} }
$$

\medskip

\textbf{Negation Introduction:}
$$
\infer[\neg_I^i]
{\dual{\ell}}
{\infer*{\bot}{[\ell]^i}}
$$
\end{calculus}
\caption{\CND's Admissible Rules for Negation}
\label{fig:CNDNegationRules}
\end{figure}

\begin{theorem}
\label{theorem:SimulationByND}
\CND simulates \UPRCL.
\end{theorem}
\begin{proof}
Given a \UPRCL derivation $\psi$ of a clause $c$ from a set of clauses $S$, we must construct a \CND derivation $\varphi$ of $c$ from $S$ (modulo variable renaming). 
We first expand $\psi$ into a tree-like proof $\psi'$: for each clause $c'$ with several children $c_1,\ldots,c_n$ (where $n > 1$), we create $n$ copies $c^1, \ldots, c^n$ of $c'$ and use each $c^k$ (for $1 \leq k \leq n$) as a parent for $c_k$. The variables in each copy are renamed to fresh variables and all substitutions in the proof are updated accordingly, in order to maintain the property that distinct clauses in $\psi'$ do not share variables (cf. Section \ref{sec:Resolution}). Now that $\psi'$ is tree-like, we may compute its \emph{global substitution} $\sigma^*$ (i.e. the composition (in topological order) of all the substitutions used in the proof)\footnote{Since we assume that distinct clauses in $\psi'$ do not share variables and $\psi'$ is tree-like, we do not need to worry about variable clashes in the composition of all substitutions. The topological order is needed because a variable $x$ introduced by a substitution $\sigma_1$ may be in the domain of another substitution $\sigma_2$ occurring below $\sigma_1$. In this case, the topologically ordered composition is $\sigma_1 \sigma_2$ (i.e. apply first $\sigma_1$ and then $\sigma_2$).}.

We now do a recursive top-down traversal of $\psi'$ and for each subderivation $\eta$ deriving a clause $c'$ from $S$ with decision literals $[\ell_1], \ldots, [\ell_n]$, we construct a corresponding subderivation $\xi$ deriving $c'~\sigma^*$ from $S$ with assumptions $[\ell_1~\sigma^*], \ldots, [\ell_n~\sigma^*]$:
\begin{itemize}
\item \emph{Base Case 1:} $\eta$ has just a leaf node containing a decision literal $[\ell]$. In this case, $\xi$ is the leaf node containing the assumption $[\ell~\sigma^*]$.

\item \emph{Base Case 2:} $\eta$ has just a leaf node containing a clause $\Gamma$. In this case, $\xi$ is:
$$
\infer[\forall_E]{\Gamma~\sigma^*}{\Gamma}
$$

\item \emph{Induction Case 1:} $\eta$ ends with a unit-propagating resolution inference, as shown below:
$$
\infer[\upr{\sigma}]
{\ell~\sigma}
{
  \infer*[\eta_1]{\ell_1}{} 
& \ldots & 
  \infer*[\eta_n]{\ell_n}{} 
& 
  \infer*[\eta']{\dual{\ell'_1} \vee \ldots \vee \dual{\ell'_n} \vee \ell}{} 
}
$$
By induction hypothesis, there are \CND derivations $\xi_1$, \ldots, $\xi_n$, $\xi'$ of, respectively, $\ell_1~\sigma^*$, \ldots, $\ell_n~\sigma^*$, $(\dual{\ell'_1} \vee \ldots \vee \dual{\ell'_n} \vee \ell)~\sigma^*$. We then construct $\xi$ by applying implication elimination $n$ times, as shown below:
$$
\infer[\imp_E]
{\ell~\sigma}
{
	\infer*[\xi_n]{\ell_n~\sigma^*}{}
	&
	\infer[\imp_E]
	{\vdots}
	{
		\infer*{\qquad}{}
	&
		\infer[\imp_E]
		{(\dual{\ell'_2} \vee \ldots \vee \dual{\ell'_n} \vee \ell)~\sigma}
		{
		  \infer*[\xi_1]{\ell_1~\sigma^*}{}
		&
		  \infer*[\xi']{(\dual{\ell'_1} \vee \ldots \vee \dual{\ell'_n} \vee \ell)~\sigma^*}{}
		}
	}
}
$$
\item \emph{Induction Case 2:} $\eta$ ends with a conflict inference. This case is analogous to the case above. But, instead of $n$ implication elimination inferences, a single negation elimination inference suffices.

\item \emph{Induction Case 3:} $\eta$ ends with a conflict-driven clause learning inference. In this case, the corresponding subproof in $\psi$ used to have the following form:
$$
\infer[\cdcl^{i}]
{(\dual{\ell_1} \sigma^1_1 \vee \ldots \vee \dual{\ell_1} \sigma^1_{m_1}) \vee \ldots \vee (\dual{\ell_n} \sigma^n_1 \vee \ldots \vee \dual{\ell_n} \sigma^n_{m_n})}
{\infer*{\bot}{\infer*[(\sigma_1^1,\ldots,\sigma_{m_1}^1)]{}{[\ell_1]^{i_1} } &  & \infer*[(\sigma_1^n,\ldots,\sigma_{m_n}^n)]{}{[\ell_n]^{i_n} } }}
$$
But due to the expansion to a tree, the subproof $\eta$ in $\psi'$ has the form shown below, where there is a copy $[\ell_k^j]$ of a decision literal $[\ell_k]$ for every path $j$ that existed from $[\ell_k]$ to $\bot$ in $\psi$. The copies have fresh variables, but are identical modulo variable renaming. For every $k$ and $j$, the substitution $\sigma_k^{j'}$ is essentially identical to $\sigma_k^j$, except for the fact that different variable names are used.
$$
\infer[\cdcl^{i}]
{(\dual{\ell_1} \sigma^{1'}_1 \vee \ldots \vee \dual{\ell_1} \sigma^{1'}_{m_1}) \vee \ldots \vee (\dual{\ell_n} \sigma^{n'}_1 \vee \ldots \vee \dual{\ell_n} \sigma^{'n}_{m_n})}
{
\infer*{\bot}
{
\infer*[\sigma_1^{1'}]{}{[\ell_1^1]^{i_1} } 
&
\ldots
&
\infer*[\sigma_{m_1}^{1'}]{}{[\ell_1^{m_1}]^{i_1} } 
&
\ldots 
& 
\infer*[\sigma_1^{n'}]{}{[\ell_n^1]^{i_n} } 
&
\ldots
&
\infer*[\sigma_{m_n}^{n'}]{}{[\ell_n^{m_n}]^{i_n} } 
}}
$$
By induction hypothesis, there is a derivation $\xi'$ with the form:
$$
\infer*{\bot}
{
\infer*[]{}{[\ell_1^1~\sigma^*] } 
&
\ldots
&
\infer*[]{}{[\ell_1^{m_1}~\sigma^*] } 
&
\ldots 
& 
\infer*[]{}{[\ell_n^1~\sigma^*] } 
&
\ldots
&
\infer*[]{}{[\ell_n^{m_n}~\sigma^*] } 
}
$$
And then a derivation $\xi$ can be constructed by applying the implication introduction rule as many times $k$ as there are assumptions $[\ell_1^1~\sigma^*], \ldots, [\ell_1^{m_1}~\sigma^*], \ldots, \ell_n^1~\sigma^*], \ldots, [\ell_n^{m_n}~\sigma^*]$ to be discharged, as depicted below:
\begin{footnotesize}
$$
\infer[\imp_I^1,\ldots,\imp_I^k]
{(\dual{\ell_1^1} \sigma^* \vee \ldots \vee \dual{\ell_1^{m_1}} \sigma^*) \vee \ldots \vee (\dual{\ell_n^1} \sigma^* \vee \ldots \vee \dual{\ell_n^{m_n}} \sigma^*)}
{
		\infer*{\qquad\bot\qquad}
		{
		\infer*[]{}{[\ell_1^1~\sigma^*]^1 } 
		% &
		% \ldots
		% &
		% \infer*[]{}{[\ell_1^{m_1}~\sigma^*] } 
		% &
		% \ldots 
		% & 
		% \infer*[]{}{[\ell_n^1~\sigma^*] } 
		&
		\ldots
		&
		\infer*[]{}{[\ell_n^{m_n}~\sigma^*]^k } 
		}
}
$$
\end{footnotesize}
Since $\sigma^*$ is the composition of all substitutions in $\psi'$, including every $\sigma_k^{j'}$, we have that $\sigma_k^{j'} \sigma^* = \sigma^*$. Therefore, the conclusion of $\xi$ is identical to:
$$((\dual{\ell_1} \sigma^{1'}_1 \vee \ldots \vee \dual{\ell_1} \sigma^{1'}_{m_1}) \vee \ldots \vee (\dual{\ell_n} \sigma^{n'}_1 \vee \ldots \vee \dual{\ell_n} \sigma^{'n}_{m_n}))~\sigma^*$$

\end{itemize}

At the end of the top-down traversal, we have a \CND proof $\varphi$ of $c~\sigma^*$ from $S$. 
Since $\sigma^*$ is the global substitution of all substitutions used in $\psi'$ and $\psi'$ derives $\kappa$, we have that $c~\sigma^* = c$. Therefore, $\varphi$ is a \CND proof of $c$ from $S$, as desired.
\end{proof}

\begin{example}
To illustrate the transformation of \UPRCL derivations into \CND derivations used in the proof of Theorem \ref{theorem:SimulationByND}, Fig.~\ref{fig:CNDDerivation} shows the \CND derivation obtained by transforming the \UPRCL derivation shown in Fig.~\ref{fig:UPRCLRefutation}.

\begin{figure*}
\centering
%\begin{tiny}
\begin{prooftree}
\AXC{$[P(a)]^2$}
		\AXC{$\neg P(a) \vee Q$} \RightLabel{$\imp_E$}
	\BIC{$Q$}
			\AXC{$[P(b)]^1$}
					\AXC{$\neg P(b) \vee \neg Q$} \RightLabel{$\imp_E$}
				\BIC{$\neg Q$} \RightLabel{$\neg_E$}
		\BIC{$\bot$} \RightLabel{$\neg_I^1$}
		\UIC{$\neg P(b)$} \RightLabel{$\imp_I^2$}
		\UIC{$\varphi_1: \neg P(a) \vee \neg P(b)$}
\end{prooftree}

\begin{prooftree}
						\AXC{$[\neg P(a)]^3$}
								\AXC{$ P(z) \vee Q $} \RightLabel{$\forall_E$}
								\UIC{$ P(a) \vee Q $} \RightLabel{$\imp_E$}
							\BIC{$Q$}
									\AXC{$[\neg P(a)]^3$}
											\AXC{$ P(y) \vee \neg Q$} \RightLabel{$\forall_E$}
											\UIC{$ P(a) \vee \neg Q $} \RightLabel{$\imp_E$}
										\BIC{$\neg Q$} \RightLabel{$\neg_E$}
								\BIC{$\bot$} \RightLabel{$\neg_I^3$}
								\UIC{$\varphi_2: P(a)$}	  
\end{prooftree}

\begin{prooftree}
		\AXC{$\varphi_2$}	
					\AXC{$\varphi_1$} \RightLabel{$\imp_E$}				            		
				\BIC{$\neg P(b)$} 
					    \AXC{$P(v) \vee \neg Q$} \RightLabel{$\forall_E$}
					    \UIC{$P(b) \vee \neg Q$} \RightLabel{$\imp_E$}
					\BIC{$\neg Q$}
							\AXC{$\varphi'_2$}
									\AXC{$\neg P(a) \vee Q$} \RightLabel{$\imp_E$}	
								\BIC{$Q$} \RightLabel{$\neg_E$}
						\BIC{$\bot$}
\end{prooftree}

where $\varphi'_2$ is a reference to a copy of $\varphi_2$.

\smallskip

%\end{tiny}
\caption{\CND Refutation Simulating the \UPRCL Refutation from Fig.~\ref{fig:UPRCLRefutation}.}
\label{fig:CNDDerivation}
\end{figure*}
\end{example}

\begin{corollary}
\UPRCL is sound.
\end{corollary}
\begin{proof}
Let $\varphi$ be an arbitrary \UPRCL proof of $c$ from $S$. Then, by Theorem \ref{theorem:SimulationByND}, there is a \CND proof of $c$ from $S$. Since the natural deduction calculus \CND is sound, $c$ is entailed by $S$. Therefore, \UPRCL is sound.
\end{proof}

% \section{Isomorphism}

% ToDo: isomorphism between UPRCL and conflict graphs.

\section{Simulation of Splitting}
\label{sec:Splitting}

Suppose that a prover refutes the set of clauses $S \cup \{ \Gamma_1 \vee \ldots \vee \Gamma_k \}$ (where the sets of variables $V_i$ of $\Gamma_i$ are mutually disjoint), by splitting it into the $k$ sets $S \cup \{\Gamma_i\}$ (for $1 \leq i \leq k$) and finding a Resolution refutation $\psi_i$ for each set $S \cup \{\Gamma_i\}$. One way to combine these proofs into a single resolution refutation of $S \cup \{ \Gamma_1 \vee \ldots \vee \Gamma_k \}$
would be to use the following recursive method:
\begin{itemize}
\item \emph{For $i = 1$:} construct $\psi'_1$ by replacing every leaf occurrence of $\Gamma_1$ in $\psi_1$ by $\Gamma_1 \vee \ldots \vee \Gamma_k$, propagating the added literals downwards and factoring the added literals when possible; then $\psi'_1$ is not a refutation, but a derivation of $\Gamma_2 \vee \ldots \vee \Gamma_k$.

\item \emph{For $i$ from $2$ to $k$:} construct $\psi'_{i+1}$ by replacing every leaf occurrence of $\Gamma_{i+1}$ in $\psi_{i+1}$ by the subproof $\psi'_i$ deriving $\Gamma_{i+1} \vee \ldots \vee \Gamma_k$; as before, propagate the added literals downwards and factor them when possible, so that $\psi'_{i+1}$ is a proof of $\Gamma_{i+2} \vee \ldots \vee \Gamma_k$, if $i+1 < k$, or $\bot$, otherwise.
\end{itemize}
However, this method is undesirable, because it requires a substantial modification of the component proofs $\psi_i$. The modified subproofs are larger (because of all the additional literals), and this may hinder readability of the proof by humans and reduce the efficiency of automatic proof checking. 

A pragmatic approach is to disregard the attempt to output a single refutation for the original problem and simply output all the separate proofs for the split problems instead. Keeping track of all splittings is important, particularly in the more general case where splitting is done recursively (i.e. where each set $S \cup \{\Gamma_i\}$ can be split further).  
This seems to be the approach taken by most automated theorem provers. Splittings performed during the proof search are recorded in the proof file in an extra-logical way, which may even violate informal semantic requirements of the TPTP proof format\footnote{TPTP's general proof format \cite{TPTP} requires that the conclusion of an inference rule be a logical consequence of its premises. This limitation prevents an easy representation of natural deduction's implication introduction rule, tableaux's $\beta$ rule or splitting. \UPRCL's conflict-driven clause learning is also affected by this limitation.}. 

In \UPRCL, splitting can be simulated in such a way that the refutations for the split sub-problems can be combined without the drawbacks that are incurred when this is done in Resolution. Suppose that $\varphi_i$ are derivations of $S \cup \{\Gamma_i\}$. Then a refutation $\varphi$ of $S \cup \{ \Gamma_1 \vee \ldots \vee \Gamma_k \}$ can be constructed by combining all the $\varphi_i$ (for $1 \leq i \leq k$) using the following recursive method:
\begin{itemize}
\item \emph{For $i = 1$:} construct $\varphi'_1$ by replacing every leaf occurrence of $\Gamma_1$ in $\varphi_1$ by the following subproof (where $\ell_i^1, \ldots, \ell_i^{n_i}$ are duals of the literals in $\Gamma_i$):
\begin{footnotesize}
$$
\infer[\upr{\varepsilon}]{\Gamma_1}{
	[\ell_2^1]^2
	&
	\ldots
	&
	[\ell_2^{n_1}]^2
	&
	\ldots
	&
	[\ell_k^1]^k
	&
	\ldots
	&
	[\ell_k^{n_k}]^k
	&
	\Gamma_1 \vee \ldots \vee \Gamma_k
}
$$
\end{footnotesize}
Then construct $\varphi'_1$ by adding a conflict-driven clause learning to the bottom of $\varphi^*_1$:

\item \emph{For $i$ from $2$ to $k$:} construct $\varphi'_i$ by replacing every leaf occurrence of $\Gamma_i$ in $\varphi_i$ by the following subproof:
$$
\infer[\cdcl^i]{\Gamma_i}
{
	\infer*[\varphi'_{i-1}]{\bot}{}
}
$$
\end{itemize}

The desired refutation $\varphi$ of $S \cup \{ \Gamma_1 \vee \ldots \vee \Gamma_k \}$ is taken to be $\varphi'_k$.

This method of simulating splitting in \UPRCL requires no internal modification of the proofs $\varphi_i$: the modified proofs $\varphi'_i$ ($2 \leq i \leq k$) are just $\varphi_i$ with a few $\cdcl$ inferences on top. Hence, there is no loss in readability, and the only overhead for automatic proof checking is caused by the extra need to check the additional $\cdcl$ inferences. If the leaf clause $\Gamma_i$ occurs only once\footnote{It may be reused many times, since $\varphi_i$ does not need to be tree-like.}, a single $\cdcl$ inference suffices, in fact. Therefore, the increase in proof size and the overhead for proof checking are negligible.

The simulation described here shows that splitting can be seen as a macro-rule that performs, for a variable-disjoint component $\Gamma_i$, batch decisions assuming the duals of all literals not in $\Gamma_i$. The first-order mechanism of decisions and conflict-drive clause learning provided by \UPRCL is, however, more general, because it allows splitting even when the components are not variable-disjoint.

\section{\UPRCL with Sequent Notation}

The proof of \UPRCL's soundness in Section \ref{sec:Soundness} demonstrates that there is a lot in common between \UPRCL and natural deduction. In the same way that natural deduction can be presented with a sequent notation, in which assumptions are listed in the antecedent of the sequent (i.e. at the left side of the turnstile symbol), \UPRCL can also be presented with a sequent notation, with decision literals kept at the antecedent. This is shown in Fig.~\ref{fig:UPRCLSequentNotation}.

\begin{figure}
\begin{calculus}
\centering

\textbf{Decision:}
$$
\infer{\ell^i \vdash [\ell]^i}{}
$$

\bigskip
\bigskip

\textbf{Initial:}
$$
\infer{\vdash c}{}
$$

if $c$ is an input clause

\bigskip
\bigskip

\textbf{Unit-Propagating Resolution:}
$$
\infer[\upr{\sigma}]{\Delta_1~\sigma, \ldots, \Delta_n~\sigma, \Delta~\sigma \vdash \ell~\sigma}{\Delta_1 \vdash \ell_1 & \ldots & \Delta_1 \vdash \ell_n & \Delta \vdash \dual{\ell'_1} \vee \ldots \vee \dual{\ell'_n} \vee \ell}
$$

where $\sigma$ is a unifier of $\ell_k$ and $\ell'_k$, for all $k \in \{1, \ldots, n \}$.

\bigskip
\bigskip

\textbf{Conflict:}
$$
\infer[\con{\sigma}]{\Delta_1~\sigma, \Delta_2~\sigma \vdash \bot}{\Delta_1 \vdash \ell & \Delta_2 \vdash \dual{\ell'}}
$$

where $\sigma$ is a unifier of $\ell_k$ and $\ell'_k$, for all $k \in \{1, \ldots, n \}$.

\bigskip
\bigskip

\textbf{Conflict-Driven Clause Learning:}
$$
\infer[\cdcl^{i}]
{\Delta \vdash (\dual{\ell_1} \sigma^1_1 \vee \ldots \vee \dual{\ell_1} \sigma^1_{m_1}) \vee \ldots \vee (\dual{\ell_n} \sigma^n_1 \vee \ldots \vee \dual{\ell_n} \sigma^n_{m_n})}
{
\Delta, \ell_1^i \sigma^1_1, \ldots, \ell_1^i \sigma^1_{m_1}, \ldots, \ell_n^i \sigma^n_1, \ldots, \ell_n^i \sigma^n_{m_n} \vdash \bot
}
$$

where $\sigma^k_j$ (for $1 \leq k \leq n$ and $1 \leq j \leq m_k$) is the composition of all substitutions used on the $j$-th path from $\ell_k$ to $\bot$.

\end{calculus}
\caption{\UPRCL with Sequent Notation}
\label{fig:UPRCLSequentNotation}
\end{figure}

With the sequent notation, it is easier to state the inference rule for conflict-driven clause learning. All the substitutions that should be applied to the literals whose duals will be part of the learned clause have already been applied to the literals in the antecedent. There is no need to look at the substitutions that have been used in the paths above. On the other hand, the presentation with sequent notation is much more redundant and bureaucratic. Whereas in the standard presentation, the use of decision literals is a powerful way to reduce the size of clauses (as in the simulation of splitting), this beneficial effect is lost in the presentation with the sequent notation, because the decision literals are carried along in the antecedents. 

For example, if we have the clause $\neg \ell_1 \vee \ldots \vee \neg \ell_n \vee \ell$, then assuming the duals of the first $n$ literals and resolving them with the clause through unit-propagation would result in the unit clause $\ell$ in the standard presentation. With sequent notation, on the other hand, we would obtain $\ell_1, \ldots, \ell_n \vdash \ell$. While this may be conceptually convenient, because it reminds us explicitly that the unit clause $\ell$ holds only under the assumptions $\ell_1, \ldots, \ell_n$, we have no reduction in size if we also count the antecedent's size. In fact, because the proof may be a non-tree-like DAG, and decision literals may be instantiated by different substitutions along different paths of the DAG, several instances of the decision literal will accumulate in the antecedent. The number of instances may be in the worst case exponential in the height of the derivation. That is one reason why the standard presentation, where the dependence of $\ell$ on assumptions and the substitutions used to instantiate the decision literals remain implicit in the derivation, is preferable. This is particularly important during proof search, in which not all inferences are useful and we do not want to apply substitutions and accumulate copies of literals unnecessarily along the derivation. We should do that only when a conflict, warranting conflict-driven clause learning, is reached.

\section{Related Work}
\label{sec:RelatedWork}

The seminal work of Baumgartner and Tinelli (\citeyear{BaumgartnerTinelli,Baumgartner}) defining the \emph{Model Evolution} (ME) procedure was probably the first lifting of DPLL to the first-order case. It was later extended with a lemma learning rule \cite{BaumgartnerTinelliLemmaLearning}, while retaining a traditional DPLL flavor (distinct from the \emph{conflict graph} approach). In model evolution, decision literals do not contain standard variables, but \emph{parameters}, which are variables with special semantics and behavior in the case of backtracking and clause learning. \UPRCL may be considered simpler, because it does not introduce the notion of parameter; however, in contrast to model evolution, for \UPRCL the problem of interpreting decision literals as a model has not been investigated yet.

More recently, Alagi and Weidenbach (\citeyear{AlagiWeidenbach}) proposed the \emph{Non-Redundant Clause Learning} (NRCL) procedure generalizing CDCL to the Bernays-Sch\"onfinkel fragment of first-order logic. They introduce the notion of \emph{blocked decisions and clauses}, which restricts the decisions that can be made and thus allows them to prove that the learned clause is non-redundant (whereas in \UPRCL they might not be). They also introduce the notion of \emph{constrained literals}, which allow more compact representation of the model. In \UPRCL, such optimizations and restrictions are intentionally avoided, in favor of a simple calculus focused on the core aspects of generalizing decisions and conflict-driven clause learning to \emph{full} first-order logic.

Bonacina, Fuhrbach and Sofronie-Stokkermans (\citeyear{Bonacina}) give a preview of a yet unpublished first-order \emph{Semantically-Guided Goal Sensitive} (SGGS) procedure inspired by CDCL. As they observe, there is a symmetry between positive and negative literals in the propositional case (i.e. in the sense that when a decision literal $\ell$ is false, $\dual{\ell}$ is true) which appears to be lost in the first-order case (i.e. because when $\ell$ is false, we cannot conclude that $\dual{\ell}$ is true; we can only conclude that $\dual{\ell}~\sigma$ is true for some $\sigma$). One of the main challenges in lifting conflict-driven clause learning to first-order lies precisely in computing and dealing with the substitution $\sigma$ when a decision literal $\ell$ leads to a conflict and a clause containing $\dual{\ell}~\sigma$ must be learned. Instead of addressing this challenge, they circumvent it by introducing the notion of \emph{uniform falsity}, according to which $\dual{\ell}$ must be true when $\ell$ is uniformly false. With this notion, clause learning is still essentially propositional and it is not triggered at every conflict (in the standard non-uniform sense of conflict). For instance, a conflict between $R(x)$ and $\neg R(b)$ does not lead to clause learning but must be repaired by revising $R(x)$ to $x \neq b \triangleright R(x)$ instead.

The variety of approaches attempting to generalize CDCL to first-order logic shows that this is not a trivial task.
The most pragmatically successful approaches so far have harnessed the power of SAT-solvers in first-order (or even higher-order) logic not by generalizing their underlying procedures but simply by employing them as \emph{black-boxes} inside a theorem prover \cite{iProver,AVATAR,Satallax}.

\section{Conclusion}
\label{sec:Conclusion}

The development of the Conflict Resolution calculus \UPRCL was initially motivated by the recent success of CDCL and by the desire to generalize its main ideas to first-order logic. However, \UPRCL can also be seen as the convergence of two ideas that actually precede CDCL by several decades. The first one is the assumption mechanism introduced by Gentzen (\citeyear{Gentzen}) in his natural deduction calculus. The second one is Robinson's generalization of the resolution rule to first-order logic through unification (\citeyear{Robinson}). \UPRCL extends resolution as natural deduction extends Hilbert-style proof systems: decision literals are essentially assumptions, and conflict driven clause learning corresponds to (several applications of) natural deduction's implication introduction rule. And whereas Robinson used unification to generalize resolution, \UPRCL uses unification to generalize conflict-driven clause learning.

From a historical perspective, what we are seeing today is similar to what happened between 1960 and 1965. In 1960, \citeauthor{DavisPutnam} defined the \emph{propositional} resolution rule, which can be regarded as an efficient machine-oriented variant of modus ponens (implication elimination). The first-order case was then handled by \emph{grounding/instantiating} the first-order problem and using the propositional resolution rule. In 1965, \citeauthor{Robinson}'s direct generalization of the resolution rule to the first-order case enabled a breakthrough in first-order automated theorem proving. Nowadays, we have a powerful \emph{propositional} conflict driven clause learning rule, which can be regarded as an efficient machine-oriented variant of implication introduction. The first-order case is being handled by essentially grounding/instantiating the problem in various ways and using the propositional rule. If history repeats itself, we might see another breakthrough when clause learning is directly lifted to the first-order case through unification, as done in the \UPRCL calculus proposed here.

A well-defined proof system is just a first step towards the development of a proof search procedure that could be implemented as an efficient theorem prover. There is much more to the efficiency of a modern SAT-solver than just the ideas of decision literals, conflict-driven clause learning and unit-propagation. SAT-solvers use restarts, strategies for selecting decision literals and data-structures that allow efficient unit-propagation, fast conflict graph analysis and fast backtracking. Adapting these proof search strategies and implementation techniques to the Conflict Resolution calculus \UPRCL is beyond the scope of this paper, but is a crucial direction for future work.

\paragraph{Acknowledgements:} Bruno is grateful to Pascal Fontaine, who supervised him during his first post-doc, providing a great opportunity for him to learn some of the essential ideas behind current SAT-solvers. Bruno is thankful to Peter Baumgartner, who shared his experience in model evolution and other related methods, when they discussed the idea of \UPRCL in May 2015. Bruno would also like to thank Hans de Nivelle and Jens Otten for discussions during the Vienna Summer of Logic about limitations of the TPTP proof format that affect the representation of natural deduction and tableau proofs. 

\bibliographystyle{abbrvnat}

\begin{thebibliography}{}
\softraggedright




% \bibitem[Sutcliffe(2015)]{CASCCompetition}
% Geoff Sutcliffe. ``Proceedings of the CADE-25 ATP System Competition (CASC-25)'' (2015).



% \bibitem[Bonacina and Plaisted(2015)]{BonacinaPlaisted}
% Maria Paola Bonacina and David A. Plaisted. ``Semantically-Guided Goal-Sensitive Reasoning''. \emph{Unpublished} (2015).

\bibitem[Alagi and Weidenbach(2015)]{AlagiWeidenbach}
Gabor Alagi and Christoph Weidenbach. ``Non-Redundant Clause Learning''. In: \emph{FroCoS} (2015), pp. 69--84.



\bibitem[Beth(1955)]{Tableaux}
Evert W. Beth. ``Semantic Entailment and Formal Derivability''. In: \emph{Mededelingen van de Koninklijke Nederlandse Akademie van Wetenschappen, Afdeling Letterkunde} 18.13 (1955), pp. 309--342.



\bibitem[Bachmair and Ganzinger(1990)]{BachmairGanzinger1990}
Leo Bachmair and Harald Ganzinger. ``Completion of First-Order Clauses with Equality by Strict Superposition (Extended Abstract)''. In: \emph{2nd International Workshop Conditional and Typed Rewriting Systems}, LNCS 516, Springer (1990), pp. 162--180.

\bibitem[Bachmair and Ganzinger(1994)]{BachmairGanzinger1994}
Leo Bachmair and Harald Ganzinger. ``Rewrite-based Equational Theorem Proving with Selection and Simplification''. In: \emph{Journal of Logic and Computation} 4.3 (1994), pp. 217--247.

\bibitem[Baumgartner(2014)]{Baumgartner}
Peter Baumgartner. ``Model Evolution Based Theorem Proving''. In: \emph{IEEE Inteligent Systems} 29(1) (2014), pp. 4--10.

\bibitem[Baumgartner and Tinelli(2003)]{BaumgartnerTinelli}
Peter Baumgartner and Cesare Tinelli. ``The Model Evolution Calculus''. In: \emph{CADE} (2003), pp. 350--364.

\bibitem[Baumgartner, Fuchs and Tinelli(2006)]{BaumgartnerTinelliLemmaLearning}
Peter Baumgartner, Alexander Fuchs and Cesare Tinelli. ``Lemma Learning in the Model Evolution Calculus''. In: \emph{LPAR} (2006), pp. 572--586.



\bibitem[Biere(2008)]{BiereTraceCheck}
Armin Biere. ``Picosat Essentials''. In: \emph{Journal on Satisfiability, Boolean Modelling and Computation (JSAT)} (2008).


\bibitem[Bonacina, Fuhrbach and Sofronie-Stokkermans(2015)]{Bonacina}
Maria Paola Bonacina, Ulrich Fuhrbach and Viorica Sofronie-Stokkermans. ``On First-Order Model-Based Reasoning''. In: \emph{Logic, Rewriting and Concurrency} (2015), pp. 181--204.


\bibitem[Brown(2012)]{Satallax}
Chad E. Brown. ``Satallax: An Automatic Higher-Order Prover''. In: \emph{IJCAR} (2012), pp. 111--117.

\bibitem[Brown(2013)]{Brown2013}
Chad E. Brown. ``Reducing Higher-Order Theorem Proving to a Sequence of SAT Problems''. In: \emph{Journal of Automated Reasoning} (2013), pp. 57--77.



\bibitem[Davis and Putnam(1960)]{DavisPutnam}
Martin Davis and Hilary Putnam. ``A Computing Procedure for Quantification Theory''. 
In: \emph{Journal of the ACM} 7 (1960), pp. 201--215.

\bibitem[Davis, Logemann and Loveland(1962)]{DLL}
Martin Davis, George Logemann and Donald Loveland. ``A Machine Program for Theorem Proving''. In: \emph{Communications of the ACM} 5(7) (1962), pp. 394--397.


\bibitem[Gentzen(1935)]{Gentzen}
Gerhard Gentzen. ``Untersuchungen \"uber das logische Schlie{\ss}en I \& II''. In: \emph{Mathematische Zeitschrift} 39.1 (1935), pp. 176--210 \& 405--431.

% \bibitem[G\"odel(1929)]{SemiDecidability}
% Kurt G\"odel. ``\"Uber die Vollst\"andigkeit des Logikkalk\"uls''. In: \emph{University of Vienna} Doctoral Dissertation (1929).

\bibitem[Korovin(2008)]{iProver}
Konstantin Korovin. ``iProver - An Instantiation-Based Theorem Prover for First-Order Logic (System Description)''. In: \emph{International Joint Conference on Automated Reasoning (IJCAR)} (2008), pp. 292--298.




\bibitem[Marques-Silva and Sakallah(1996)]{GRASP}
Joao Marques-Silva and K.A. Sakallah. ``GRASP: A New Search Algorithm for Satisfiability''. In: \emph{International Conference on Computer-Aided Design} (1996), pp. 220 -- 227.

\bibitem[Marques-Silva et al.(2008)]{CDCL}
Joao Marques-Silva, Ines Lynce and Sharad Malik. ``Conflict-Driven Clause Learning SAT Solvers''. In: \emph{Handbook of Satisfiability} (2008), pp. 127 -- 149.



\bibitem[McCharen, Overbeek and Wos(1976)]{UnitResultingResolution}
J. McCharen, R. Overbeek and L. Wos. ``Complexity and Related Enhancements for Automated Theorem-Proving Programs''. In: \emph{Computers and Mathematics with Applications} 2 (1976), pp. 1--16.

\bibitem[McCune(2006)]{Prover9Manual}
W. McCune. ``Prover9Manual'' (2006).

\bibitem[Riazanov and Voronkov(2002)]{Vampire}
Alexandre Riazanov and Andrei Voronkov. ``The Design and Implementation of VAMPIRE''. In: \emph{AI Communications} 15(2-3)(2002), pp. 91--110.



\bibitem[Robinson(1960)]{Robinson}
John Alan Robinson. ``A Machine-Oriented Logic Based on the Resolution Principle''. 
In: \emph{Journal of the ACM} 12.1 (1965), pp. 23--41.

\bibitem[Robinson and Wos(1969)]{Paramodulation}
George Robinson and Larry Wos. ``Paramodulation and Theorem-Proving in First-Order Thories with Equality''. In: \emph{Machine Intelligence} 4 (1969), pp. 135--150.



\bibitem[Schultz(2013)]{E}
Stephan Schultz. ``System Description: E 1.8''. In: \emph{LPAR} (2013), pp. 735--743.

\bibitem[Sutcliffe(2009)]{TPTP}
Geoff Sutcliffe. ``The TPTP Problem Library and Associated Infrastructure: 
                    The FOF and CNF Parts, v3.5.0''. In: \emph{Journal of Automated Reasoning} 43.4 (2009), pp. 337--362.

\bibitem[Voronkov(2014)]{AVATAR}
Andrei Voronkov. ``AVATAR: The Architecture for First-Order Theorem Provers''. In: \emph{CAV} (2014), pp. 696--710.



\bibitem[Waldmann(2015)]{WaldmannEPSSuperposition}
Uwe Waldmann. ``Superposition''. In: \emph{Encyclopedia of Proof Systems} (2015).

\bibitem[Waldmann(2015)]{WaldmannEPSSaturation}
Uwe Waldmann. ``Saturation with Redundancy''. In: \emph{Encyclopedia of Proof Systems} (2015).

\bibitem[Weidenbach(2001)]{WeidenbachSplitting}
Christoph Weidenbach. ``Combining Superposition, Sorts and Splitting''. In: \emph{Handbook of Automated Reasoning} (2001), pp. 1965--2013.

\bibitem[Weidenbach(2002)]{SPASSManual}
Christoph Weidenbach. ``The Theory of SPASS Version 2.0''. In: \emph{SPASS 2.0 Documentation}.

\bibitem[Weidenbach et al.(2009)]{SPASS}
Christoph Weidenbach, Dilyana Dimova, Arnaud Fietzke, Rohit Kumar, Martin Suda, Patrick Wischnewski. ``SPASS Version 3.5''. In: \emph{CADE} (2009), pp. 140--145.

\bibitem[Wetzler, Heule and Hunt Jr.(2014)]{DRATTrim}
Nathan Wetzler, Marijn Heule and Warren A. Hunt Jr. ``DRAT-trim: Efficient Checking and Trimming Using Expressive Clausal Proofs''. In: \emph{SAT} (2014), pp. 422--429.

\bibitem[Zhang et al.(2001)]{FirstUIP}
Lintao Zhang, Conor F. Madigan, Matthew H. Moskewicz, Sharad Malik. ``Efficient Conflict Driven Learning in a Boolean Satisfiability Solver''. In: \emph{International Conference on Computer-Aided Design} (2001), pp. 279--285.


\end{thebibliography}

% The bibliography should be embedded for final submission.

%\clearpage

\section*{Appendix - A Standard Non-Clausal Classical Natural Deduction Calculus}

A standard natural deduction calculus for minimal quantified logic extended with a classical rule for double negation elimination is shown in Fig.~\ref{fig:ND}.

%ToDo: do we need ex falso?

\begin{figure}[!h]
\begin{calculus}
\centering

\textbf{Implication Elimination (Modus Ponens):}
$$
\infer[\imp_E]{B}{A & A \imp B}
$$

\medskip

\textbf{Implication Introduction:}
$$
\infer[\imp_I^i]
{A \imp B}
{\infer*{B}{[A]^i}}
$$

\medskip

\textbf{Universal Quantification Elimination:}
$$
\infer[\forall_E]{A \{ \sub{x}{t} \} }{A}
$$

\medskip

\textbf{Universal Quantification Introduction:}
$$
\infer[\forall_I]
{A}
{A \{ \sub{x}{\alpha} \} }
$$

$\alpha$ must be an \emph{eigen-variable}: \\ it should occur neither in $\Gamma$ nor in any undischarged assumption.

\medskip

\textbf{Double Negation Elimination:}
$$
\infer[\dot{\neg}\dot{\neg}_E]
{A}
{(A \imp \bot) \imp \bot}
$$
\end{calculus}

\caption{A Non-Clausal Natural Deduction Calculus}
\label{fig:ND}
\end{figure}

% \section{Quasimodels}

% ToDo: this section is very drafty. It is the beginning of an attempt to give semantic meaning to that vague notion of first-order conflict graph that we have been discussing. The idea is that, even though the literals in the graph can be contradictory, if the graph has been grown enough (according to a yet undefined notion of maximality), we can fix the model by removing some contradictory literals.

% \begin{definition}
% A decision literal or propagated literal occurring in a node $\eta$ of a {\UPRCL}-derivation $\psi$ is \emph{alive} if and only if there is a sink node $\eta'$ not holding $\bot$ below $\eta$.
% \end{definition}

% \begin{definition}
% The \emph{quasimodel} $M_{\psi}$ of a {\UPRCL}-derivation $\psi$ is the set of live literals in $\psi$.
% \end{definition}

% A quasimodel may be incomplete (i.e. not satisfying all input clauses) or inconsistent (i.e. it may contain contradictory literals).

% \begin{theorem} (Conjecture)
% If $M$ is a maximal complete quasimodel, then there is a consistent subset $M'$ which is model of the input clauses.
% \end{theorem}

\end{document}